\documentclass[twocolumn,floats,aps]{revtex4}
\usepackage{times,amsmath,graphics,psfrag,latexsym,pstricks}

\begin{document}  
\title{Packing Fractions and Maximum Angles of Stability of Granular Materials}  
\author{J. Olson, M. Priester, J. Luo, S. Chopra and R.J. Zieve}
\affiliation{Physics Department, University of California at Davis}  
\begin{abstract}  
In two-dimensional rotating drum experiments, we find two separate influences of
the packing fraction of a granular heap on its stability.  For a fixed grain
shape, the stability increases with packing fraction.  However, in determining
the relative stability of different  grain shapes, those with the {\em lowest}
average packing fractions tend to form the most stable heaps.  We also show that
only the configuration close to the surface of the pile figures prominently.  
\end{abstract}
\maketitle

Granular materials are studied in materials science, geology, physics, and
engineering \cite{Duran}.  Their unusual dynamical behavior has attracted much recent
attention \cite{Kudrolli, Rajchenbach, Behringer97}, and even the more staid static issues generate steady
interest \cite{Edwards, Bideau}.  One question is how much of granular behavior can be
derived from purely geometrical considerations.  In practical materials,
particle deformation and surface effects such as cohesion and agglomeration
also play major roles, masking the influence of geometry. 
Furthermore, studying grain shape is a difficult endeavor.  The experimental
challenge is in creating uniform but non-spherical shapes.  Recent efforts
include two-dimensional studies of regular pentagons \cite{Cantelaube, Troadec} and a
three-dimensional experiment using M\&M's \cite{Donev}.  On the theoretical
side, the problem is how to treat the interaction of grains as they move past
each other.

Work linking geometry to behaviors other than packing is yet more scarce.
One rare attempt involves predicting avalanches in a granular pile
\cite{Albert}, a question at the border between statics and dynamics.
Granular piles, such as sandpiles, exhibit characteristic angles for
their free surfaces.  One is the repose angle, below which the pile
is stable.  If the surface becomes steeper than the repose angle,
for example if the pile is tilted or if new grains are added, then the
heap may undergo an avalanche in which grains all along the slope move,
resulting in a lower angle.  In practice, avalanches do not begin until
the angle exceeds what is known as the maximum angle of stability, which
is typically several degrees larger than the repose angle.  Once started,
an avalanche continues until the pile surface is again less steep than
the repose angle.  
Albert et al. \cite{Albert} derive a maximum
angle of stability from local geometry, beginning with a regular tetrahedron
of spheres.
In addition to safety issues, understanding avalanches
is important in fields such as geology and soil mechanics, where granular
matter can flow along inclined surfaces.

Both theoretical and experimental work relate the packing fraction
of a heap to its stability, for the special case of spherical grains
\cite{Brown, Allen, Evesque93}.  The experiments involve packing spheres under
pressure to achieve different initial packing fractions, then tilting the
heap and noting the angle of the first avalanche.
As the spheres are packed more tightly, the maximum angle of stability
increases.  In both these measurements, the packing fraction is known
only for the initial, artificially constructed arrangement.  They do not
test whether packing fraction also affects stability in the configurations
that actually occur after an avalanche.  One set of these experiments
\cite{Brown} uses not only smooth spheres but also
``rough" spheres and angular grains.  The two very different non-spherical
shapes sustain similar maximum angles, higher than that of smooth spheres.

Here we revisit the role of density, extending our study to naturally
occurring configurations and to non-spherical grains.  We work in
two dimensions, which makes visualizing an entire configuration much
easier than for a three dimensional heap.  Our grains are composed of
spherical ball bearings, welded together in clusters of up to 9 balls.
The balls in each cluster are part of a two-dimensional triangular
lattice.  Working with sphere clusters has various advantages. The
maximum possible density is always that of a triangular lattice of
the component spheres.  Spheres minimize friction and blocking effects
as the shapes move past each other.  Finally, our system lends itself
to comparison with computer experiments, since checking for overlaps,
a challenging part of typical simulations, is trivial for sphere clusters.

We find that packing fraction indicates pile stability for both
spherical and non-spherical shapes, but with a twist.  When comparing
piles composed of different grains, {\em low} average packing fraction
indicates stability.  After presenting our experimental results, we
offer an explanation for this behavior and other observations about the
influence of grain shape.

As described elsewhere \cite{Ivan}, we weld together $\frac18$-inch diameter
carbon steel ball bearings to make dimers; trimers (three balls in a straight
line); triangles of three or six balls; diamonds of four, six, or nine balls;
trapezoids of five or seven balls; and hexagons of seven balls.  The left
column of Table \ref{t:critangle} illustrates these shapes.

\begin{figure}[htb]
\scalebox{0.9}{\includegraphics{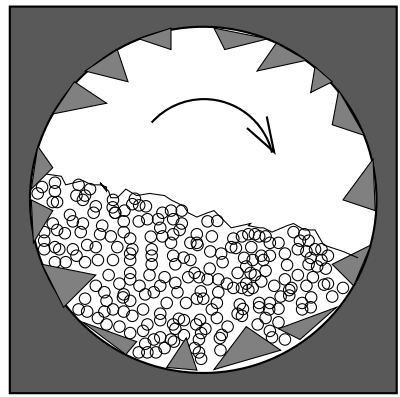}}
\caption{Tumbler for two-dimensional granular heap.  The triangles transform the
container boundary from a circle to an irregular shape.}
\label{f:tumbler}
\end{figure}

Our tumbler, shown schematically in Figure \ref{f:tumbler}, is a sheet
of aluminum, $\frac18$" thick with a circular hole 14 inches in diameter
cut from its center.  The aluminum is sandwiched between two $\frac12$"
thick sheets of Plexiglas, which constrain the balls to move in a
single layer.  The plane of the tumbler is vertical.  A central axle
attaches the tumbler to a stepper motor which controls the rotation rate.
For all measurements here the rotation rate is about 500 $\mu$Hz,
or one full turn in about thirty minutes.  We use this slow speed so
that the rotation of the container itself during an avalanche remains
negligible, and the avalanches are discrete events.  To prevent the
balls from sliding along the wall of the container during rotation,
a thin strip of rubber is glued to the inner edge of the aluminum.
In some measurements, as described below, an irregular boundary was
created by attaching aluminum triangles along the edge of the hole.

For each shape we use a total mass of 365.6$\pm0.3$ g, so that the grains fill
a similar portion of the tumbler.  We rotate the tumbler for about 30 minutes,
which generally yields 20 to 30 discrete avalanches.  The entire rotation is
recorded with a digital video camera. Afterwards the avalanches  are identified
by eye, and the video frames immediately before and after each avalanche are
uploaded to a computer.  

To simplify the image processing, we use a solid red background inside
the tumbler and solid white around the outside.  These colors allow a
computer program to identify easily the region occupied by the heap.
Since we know both the total mass of the grains and the mass per grain,
we can convert the area of the heap to a packing fraction.  We also fit
a line to the free surface of the heap and use it to define the angle
from horizontal of the surface.  In this way we extract the
angle and packing fraction for each image.

On removing the shapes from the tumbler after a measurement, we sort and
count any broken grains.  The maximum angle of stability is sensitive to
broken shapes; so if over 10\% of the original pieces break during
a measurement, we discard the data and run that shape again.  In most cases
breakage is less than 3\% of the original shapes.

\begin{figure}[hbt]
\psfrag{sin}{\pspicture(-.3,-.2)(.4,.2)
\pscircle(0,0){.12}
\endpspicture }
\psfrag{db}{\pspicture(-.4,-.2)(.4,.2)
\pscircle(-.16,0){.16}
\pscircle(.16,0){.16}
\endpspicture }
\psfrag{hex}{\pspicture(-.4,-.2)(.4,.2)
\pscircle(-.32,0){.16}
\pscircle(0,0){.16}
\pscircle(.32,0){.16}
\pscircle(-.16,.2771){.16}
\pscircle(.16,.2771){.16}
\pscircle(-.16,-.2771){.16}
\pscircle(.16,-.2771){.16}
\endpspicture }
\psfrag{stri}{\pspicture(-.4,-.1)(.4,.2)
\pscircle(-.16,0){.16}
\pscircle(.16,0){.16}
\pscircle(0,.2771){.16}
\endpspicture  }
\psfrag{btri}{\pspicture(-.4,-.25)(.4,.2)
\pscircle(-.16,0){.16}
\pscircle(.16,0){.16}
\pscircle(0,.2771){.16}
\pscircle(-.32,-.2771){.16}
\pscircle(0,-.2771){.16}
\pscircle(.32,-.2771){.16}
\endpspicture }
\psfrag{diam}{\pspicture(-.4,-.1)(.4,.2)
\pscircle(-.16,0){.16}
\pscircle(.16,0){.16}
\pscircle(0,.2771){.16}
\pscircle(.32,.2771){.16}
\endpspicture }
\psfrag{trip}{\pspicture(-.4,-.2)(.4,.2)
\pscircle(-.32,0){.16}
\pscircle(.32,0){.16}
\pscircle(0,0){.16}
\endpspicture }
\psfrag{trap}{\pspicture(-.4,-.1)(.4,.2)
\pscircle(-.32,0){.16}
\pscircle(0,0){.16}
\pscircle(.32,0){.16}
\pscircle(-.16,.2771){.16}
\pscircle(.16,.2771){.16}
\endpspicture }
\scalebox{.35}{\includegraphics{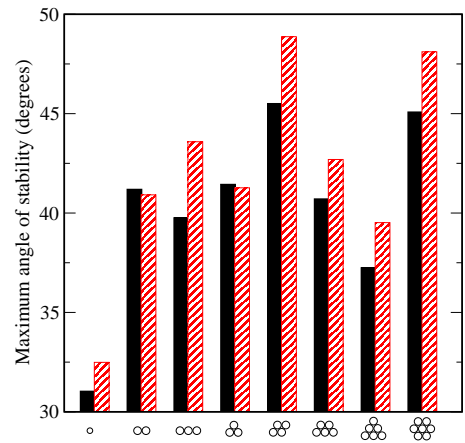}}
\caption{Average angle for avalanche onset for eight different shapes.  Black:
container with circular boundary.  Red: container with irregular boundary.
The shapes used are indicated along the horizontal axis, arranged in order
of size.}
\label{f:critangle}
\end{figure}

The black bars of Figure \ref{f:critangle} show the average angle just
before an avalanche for several different shapes.  Standard errors are
typically one degree, and never more than 1.5 degrees, so the variations
among shapes are significant.  Two shapes, diamonds and hexagons, form
significantly more stable heaps than the other shapes.  

These angles are very different from the results of our previous work in a
rectangular container, which allows a single crystalline region \cite{Cynthia}.
To test that boundary effects do not dominate the variations we find
among shapes, we inserted 16
triangles in an irregular pattern along the surface of the circle and repeated
the measurements.  The results, shown as red bars in Figure \ref{f:critangle},
are qualitatively the same.  Diamonds and hexagons remain much more stable than
the other shapes.

The quantitative differences between the containers could be an effect of the
boundary surface.  Another contribution may come from small differences in the
numbers of broken shapes in the two sets of measurements.  

In the irregular container, we extended the measurements to several additional
shapes.  All data from this setup are shown in Table \ref{t:critangle}.
There is no clear pattern of how geometry affects critical angle,
beyond the general observation that larger shapes support higher angles.
Given the scalloped boundaries of our shapes, which allow some interlocking,
this seems natural.

\begin{table}[htb]
\caption{Average maximum angle of stability ($\theta_m$) and repose angle
($\theta_r$) for eleven shapes in a container with irregular boundary. 
Standard errors are also given.  Shapes are ordered from smallest to largest.}
\label{t:critangle}
\begin{tabular}{l|llll}
& $\theta_m$ & $\sigma(\theta_m)$ & $\theta_r$ & $\sigma(\theta_r)$\\
\hline
\pspicture(-.4,-.1)(.4,.2)
\qdisk(0,0){.05}
\endpspicture &
33.8 \hspace*{.1in}  &0.7\hspace*{.1in}  & 26.3\hspace*{.1in} &0.6\\
\pspicture(-.4,-.1)(.4,.2)
\qdisk(-.07,0){.05}
\qdisk(.07,0){.05}
\endpspicture &
42.0   &0.6& 32.9&0.5\\
\pspicture(-.4,-.1)(.4,.2)
\qdisk(-.14,0){.05}
\qdisk(0,0){.05}
\qdisk(.14,0){.05}
\endpspicture &
43.8   &1.0& 34.5&0.8\\
\pspicture(-.4,-.05)(.14,.175)
\qdisk(-.07,0){.05}
\qdisk(.07,0){.05}
\qdisk(0,.12124){.05}
\endpspicture & 
41.2	&0.5& 31.3&0.3\\
\pspicture(-.35,-.05)(.4,.175)
\qdisk(-.07,0){.05}
\qdisk(.07,0){.05}
\qdisk(0,.12124){.05}
\qdisk(.14,.12124){.05}
\endpspicture & 
49.1	&1.0& 38.2&0.9\\
\pspicture(-.4,-.05)(.4,.175)
\qdisk(-.14,0){.05}
\qdisk(0,0){.05}
\qdisk(.14,0){.05}
\qdisk(-.07,.12124){.05}
\qdisk(.07,.12124){.05}
\endpspicture & 
44.9	&0.8& 33.6&0.7\\
\pspicture(-.4,-.18)(.4,.24)
\qdisk(-.07,0){.05}
\qdisk(.07,0){.05}
\qdisk(0,.12124){.05}
\qdisk(-.14,-.12124){.05}
\qdisk(0,-.12124){.05}
\qdisk(.14,-.12124){.05}
\endpspicture & 
39.5	&0.4& 29.3&0.5\\
\pspicture(-.35,-.05)(.27,.175)
\qdisk(-.14,0){.05}
\qdisk(0,0){.05}
\qdisk(.14,0){.05}
\qdisk(-.07,.12124){.05}
\qdisk(.07,.12124){.05}
\qdisk(.21,.12124){.05}
\endpspicture & 
51.8	&1.0& 37.8&0.9\\
\pspicture(-.4,-.18)(.4,.24)
\qdisk(-.14,0){.05}
\qdisk(0,0){.05}
\qdisk(.14,0){.05}
\qdisk(-.07,.12124){.05}
\qdisk(.07,.12124){.05}
\qdisk(-.07,-.12124){.05}
\qdisk(.07,-.12124){.05}
\endpspicture & 
48.7	&1.3& 35.0&1.0\\
\pspicture(-.3,-.05)(.4,.175)
\qdisk(-.14,0){.05}
\qdisk(0,0){.05}
\qdisk(.14,0){.05}
\qdisk(-.07,.12124){.05}
\qdisk(.07,.12124){.05}
\qdisk(.28,0){.05}
\qdisk(.21,.12124){.05}
\endpspicture & 
50.6	&1.1& 38.7&0.7\\
\pspicture(-.33,-.18)(.4,.24)
\qdisk(-.14,0){.05}
\qdisk(0,0){.05}
\qdisk(.14,0){.05}
\qdisk(-.07,.12124){.05}
\qdisk(.07,.12124){.05}
\qdisk(.21,.12124){.05}
\qdisk(-.21,-.12124){.05}
\qdisk(-.07,-.12124){.05}
\qdisk(.07,-.12124){.05}
\endpspicture & 
50.9	&1.5& 36.6&0.9
\end{tabular}
\end{table}

\begin{figure}[hbt]
\scalebox{.35}{\includegraphics{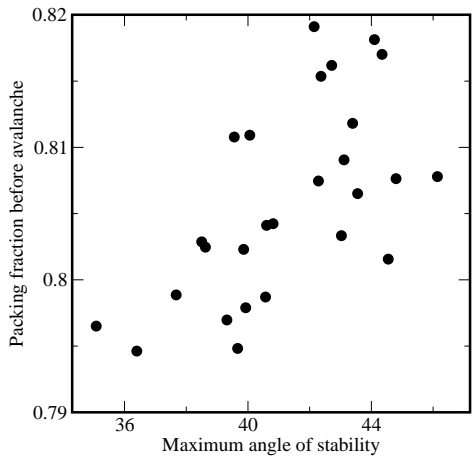}}
\caption{Packing fraction and critical angle for the individual
avalanches of the small triangle shapes (3 balls).  The correlation coefficient
is 0.626.}
\label{f:densangst}
\end{figure}

We also characterize each configuration by its packing fraction.   Figure
\ref{f:densangst} shows packing fraction and maximum angle attained for each
avalanche with the small triangles.  As in the earlier work with compacted
media in three dimensions \cite{Brown, Allen, Evesque93}, the two properties
are positively correlated. This also agrees with the observation that
interlocking and jamming, which can increase the stability angle, occur more
easily at higher packing fractions \cite{Cynthia}.  Strikingly, eight of the
ten other shapes exhibit a similar correlation between the packing fraction and
the angle of the ensuing avalanche, with correlation coefficients between 0.4
and 0.71.  (The remaining shapes, the large triangles and large trapezoids,
show no significant correlation between the two.)  Thus we confirm
experimentally that a connection between pile density and maximum angle of
stability exists for a variety of shapes, although it may not be
universal.

Even more importantly, our measurements relate these properties for naturally
occurring configurations rather than for the artificially compacted initial
arrangements of the previous work.  Finding similar results for packings created so differently and
in both two and three dimensions raises the hope of a general principle
applying to a wide variety of configurations.  Furthermore, in many practical
situations, a pile is in fact formed from previous avalanches.  Our work also
highlights the importance of packing fraction in determining stability, since
we detect effects of pile density despite its much smaller variation for our
samples than in the compressed piles.

\begin{figure}[hbt]
\scalebox{.35}{\includegraphics{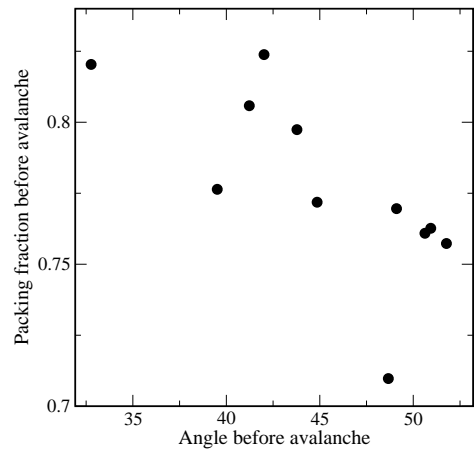}}
\caption{Average packing fraction and critical
angle for the eleven different shapes.  Each filled circle represents one
shape.}
\label{f:densangall}
\end{figure}

In addition to connecting pile density and stability angle for the individual
avalanches of each shape, we compare the behavior among shapes. Here a far less
intuitive influence of packing fraction appears. Figure \ref{f:densangall} again
plots packing fraction against maximum angle, but here each point represents an
average over all avalanches for a single shape.  Now the correlation is actually in
the {\em opposite} direction from that of Figure \ref{f:densangst}: high
packing fractions tend to yield {\em low} critical angles.

The apparently opposite influence of packing fraction within a shape and
across different shapes may be understood by considering how the pile is
formed.  From our measurements on individual shapes, we know that tightly
packed configurations are more stable.  During an avalanche, a necessary
condition for flow to cease is a stable instantaneous arrangement of
grains.  Furthermore, grain motion always requires expansion so that
the particles have room to move \cite{Reynolds}.  As the avalanches
stops, the packing fraction again increases.  Thus the moving grains
pass through configurations with a range of packing fractions.  Many of
the arrangements with the lowest packing fractions are unstable, or at
least not stable enough to absorb the momentum of the moving grains.
As the packing fraction increases, the arrangements are more likely
to be stable and allow the avalanche to stop.  If a shape has a low
average packing fraction, its avalanches must stop relatively early,
and the shape is likely to support a wide range of stable configurations.

Rotation is another means of sampling different grain arrangements, since
the stability of any configuration depends on the direction of gravity.
If more arrangements than usual are stable for a particular grain shape,
then the new configurations reached by tilting a heap of such grains
are more likely to be stable.  On average the heap will take longer to
reach an unstable configuration and trigger an avalanche.  The negative
correlation we observe between packing fraction and maximum angle of
stability comes about because low-density configurations are most likely
to occur with shapes of generally high stability. This logic implies
that particularly high critical angles should be found for tightly packed
grains with a low random loose packing value.

\begin{table}[htb]
\caption{Correlations among packing fractions before ($\rho_i$) and
after ($\rho_f$) an avalanche, maximum angle of stability ($\theta_m$),
angle of repose ($\theta_r$), and change in angle during an avalanche
($\Delta\theta$).  Values are the averages of the correlation coefficients
calculated for all 11 shapes.}
\label{t:correlations}
\begin{tabular}{l|cccc}
& $\rho_i$ & \hspace*{.2in}$\theta_r$ & $\rho_f$ & \hspace*{.2in}$\Delta\theta$\hspace*{.1in}\\
\hline
$\theta_m$ & 0.475 & -0.269 & 0.032 & 0.841\\
$\rho_i$ & & -0.105 & 0.613 & 0.399\\
$\theta_r$ & & & 0.290 & -0.728\\
$\rho_f$ & & & & -0.138
\end{tabular}
\end{table}

Monitoring successive avalanches allows us to 
study other correlations among the configurations as well.
Table \ref{t:correlations} deals with five variables: packing fraction
and angle before and after each avalanche, and the change in angle.
The table shows the correlation between each pair of quantities, averaged
over all eleven shapes.  There is no significant correlation between initial
packing fraction and repose angle, or between final packing fraction and
maximum angle of stability.  We also find directly that the angles of
successive avalanches are uncorrelated.  Previous experiments found a similar
lack of correlation in the size of successive avalanches \cite{Evesque91},
suggesting that a single avalanche completely resets the system memory. 

On the other hand, the packing fractions of successive avalanches (or
equivalently, the packing fractions before and after a single avalanche)
are highly correlated.  This is hardly surprising, given that any single
avalanche leaves about half of the ball bearings completely unaffected.
In addition to the stationary regions, there are typically several
large clumps within the pile that rotate slightly during the avalanche
but have not change in their internal arrangements.  With this in
mind, the correlation is if anything lower than expected.
The implication is that the density varies more
strongly in the top few layers, which reconfigure completely during an
avalanche, than in the rest of the heap.

Combining these observations, we see that significant packing fraction
correlations between consecutive avalanches arise from the stationary
lower layers, but that successive avalanches show no correlation
in stability angle.  Consequently, the correlation between packing
fraction and maximum angle of stability must depend {\em only} on the
packing fraction in the upper, mobile layers.  The heap density within
these layers, which we cannot easily calculate, would likely show a much
stronger relationship with the stability angle than the overall density
as in Figure \ref{f:densangst}.

Finally, there is a small but consistent negative correlation between
critical angle and repose angle. This may happen because more momentum
builds up during an avalanche that begins on a steep slope, enabling the
avalanche to continue longer.  The one exception is the single balls,
which have positive correlation between critical and repose angles as
well as much weaker connections between avalanche size and the initial
and final angles.

In conclusion, we show that packing fraction plays a dual role in predicting
the stability of a two-dimensional heap.  For a given grain shape, denser
packings are generally more stable; and we have extended this result beyond the
previous measurements on artificially packed spheres in three dimensions. 
However, when comparing different shapes, those with the {\em lowest} packing
fractions have the highest maximum stability angles on average.  We also find
that only the packing fraction of the top few layers of the heap figures
strongly affects stability, so the correlations would likely be much cleaner if
a packing fraction calculation could be confined to this region.

It would be interesting to test how widely the inverse correlation between
maximum stability angle and packing fraction holds.  The shapes we used
are all clusters of spheres, confined to two dimensions.  Experiments or
simulations could also be done on shapes such as ellipses or polygons,
and extended to three dimensions.

We are also expanding our work to heaps containing a mixture of grain shapes,
which drastically enlarges the phase space for measurements.
In the course of the present work, we observed that a small fraction of
broken shapes can significantly change the critical angle.  We expect tests
of shape mixtures to help explain why.

We thank C. Olson and C. Reichhardt for discussions.
This work was supported by the National Science Foundation under DMR-9733898.

\end{document}